\documentclass[preprint,12pt]{elsarticle}
\usepackage{booktabs}
\usepackage{array}

\usepackage{amssymb}

\journal{Vacuum}

\begin{document}

\begin{frontmatter}



\title{Lithium fluoride (LiF) target preparation for nuclear physics experiment}


\author[inst1,inst2]{Lalit Kumar Sahoo}

\affiliation[inst1]{Saha Institute of Nuclear Physics,
            addressline={Bidhan nagar}, 
            city={Kolkata},
            postcode={700064}, 
    country={India}}
    \affiliation[inst2]{Homi Bhaba National Institute,
            addressline={Anushaktinagar}, 
            city={Mumbai},
            postcode={400094}, 
           country={India}}

\author[inst1,inst2]{Ashok Kumar Mondal}
\author[inst1,inst2]{Dipali Basak}
\author[inst1]{Chinmay Basu}
\author[inst1]{Suraj Kumar Karan}

\begin{abstract}
The LiF target preparation on self-supporting Ag backing (LiF/Ag) is discussed in a detailed manner using vacuum evaporation process. The target thickness is measured using the energy loss of three line alpha source. 183.74 $\mu$g/cm$^2$ thickness of LiF is achieved through the evaporation process. Good uniformity of targets is observed. Non-uniformity in targets is found within 6 $\%$. The XPS analysis confirms the presence of both the F and Li atoms on sample surface.  

\end{abstract}





\end{frontmatter}



\section{Introduction}
\label{sec:sample1}
Targets in nuclear physics are the systems that are subjected to the bombardment of accelerated charged particles such as protons, neutrons, or heavy ions.  Nuclear targets are a key factor in the accomplishment of nuclear reactions.The use of high quality targets is essential for successful nuclear interactions. Cross-sections resulting from the nuclear interaction  are highly dependent on the quality of targets used in the experiment. The importance of target quality increases particularly in astrophysical reactions where cross sections are in order of ~nb or pb. Lithium fluoride (LiF) on self supported Ag foils (LiF/Ag) is one of the crucial target for many reactions like $^{19}$F(p,$\alpha$), $^{19}$F(p,$\gamma$), $^{7}$Li(p,$\alpha$), $^{7}$Li(p,n), $^{19}$F(p,n) etc \cite{1,2,3,4}. Preparation of a thin target is also an essential requirement for many reactions where energy loss through target would be less. LiF is a good alternative to prepare thin targets as compared to metallic lithium or fluorine. LiF has been preferred for generation of mono-energetic neutrons \cite{2,5} which can be efficiently used as neutron beam. Both the target requirement of Li and F are satisfied with the deposition of LiF compound on Ag backing. However, the detailed and efficient description of LiF target preparation was not discussed in great detail. Smail Damache et al.\cite{6} has described the LiF target preparation in some details . In this paper, we presented a detailed description of the smooth preparation of lithium fluoride (LiF) target on self-supporting silver backing by vacuum evaporation method with analysis of prepared targets.
\begin{figure}
\begin{center}
\begin{tabular}{cc}
\includegraphics[width=65mm]{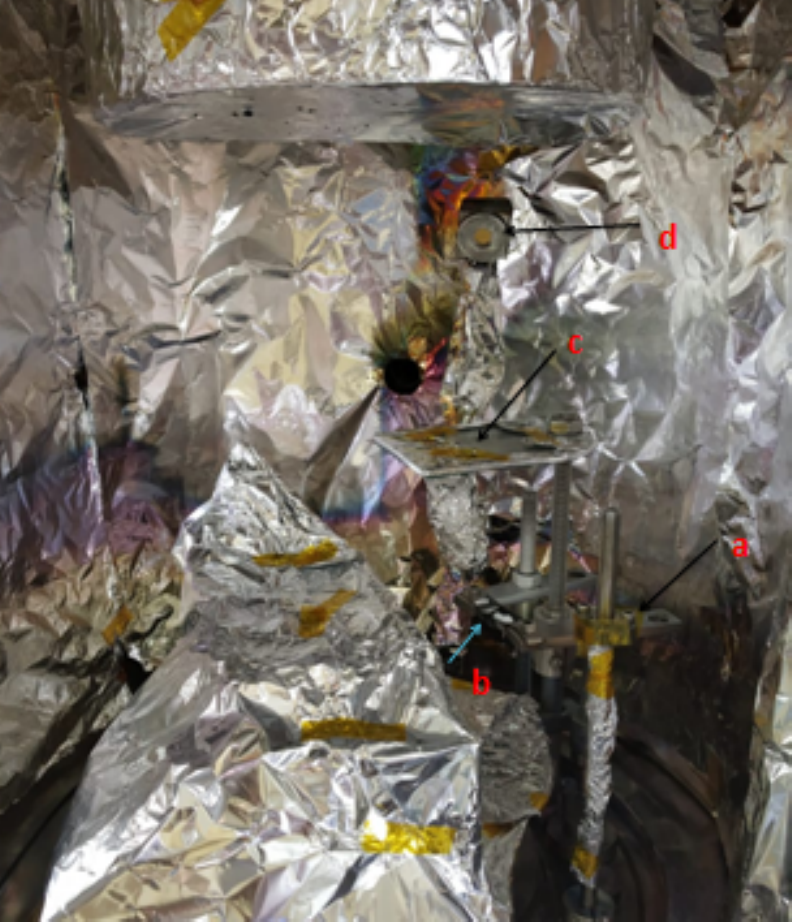}
\includegraphics[width=65mm]{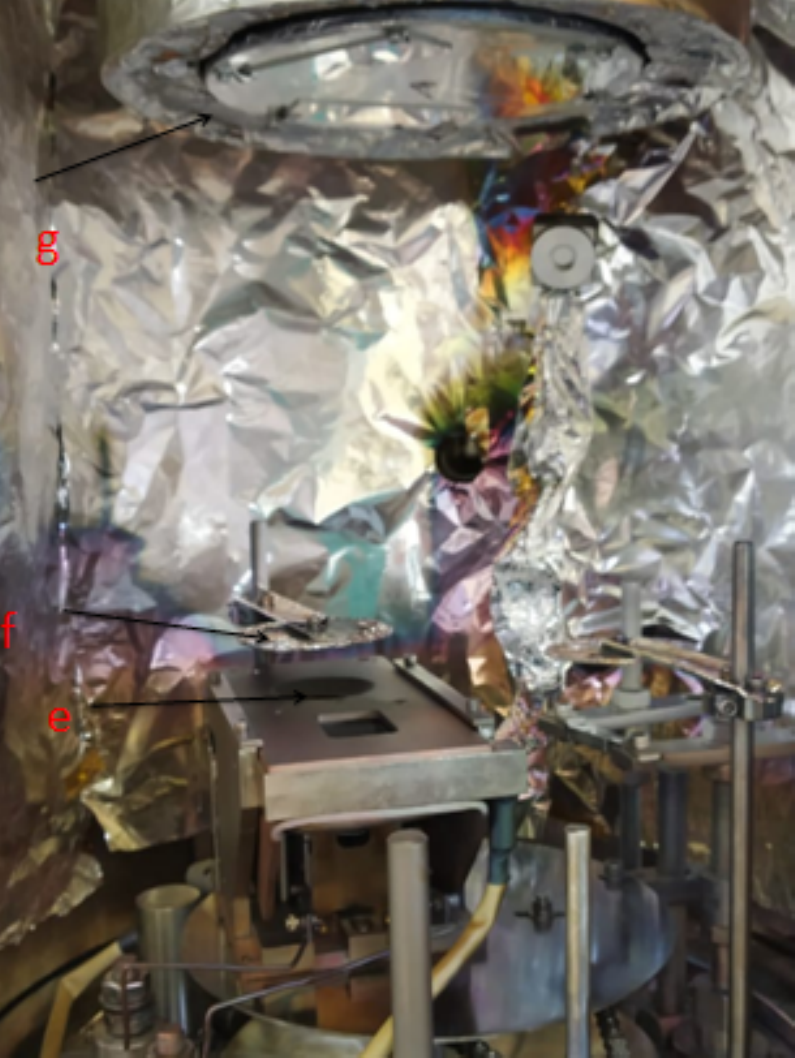}

\end{tabular}

\caption{(left) Set-up for thermal evaporation (a)Copper electrode for resistive heating (b) Mounted boat for placing sample (c)Substrate holder (d) Quartz. (Right) Setup for e- beam evaporation in same chamber (e) Crucible for placing sample (f) e- beam shutter (g) Substrate holder with rotating plate.}
\label{fig:1} 
\end{center}      
\end{figure}
\begin{figure}
\begin{center}

\includegraphics[width=75mm]{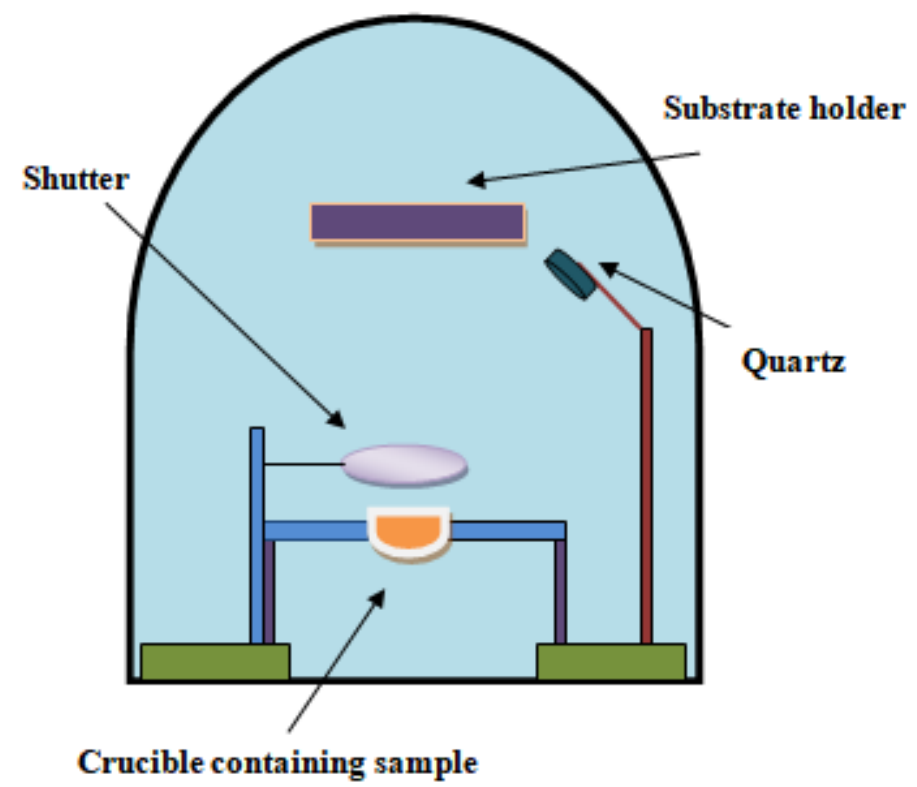}

\caption{Schematic View of Vacuum Chamber}
\label{fig:2} 
\end{center}      
\end{figure}
\section{Deposition Set-up}
\label{sec:sample2}
The deposition setup is divided into two parts according to the mode of deposition. One part is for thermal evaporation and  second one is for e$^-$-beam evaporation. Figure 1 (left) shows the thermal evaporation part. It consists of a copper electrode and a boat that hold the sample. The Copper electrode is used for resistive heating purposes in the application of current. One substrate holder is placed above it as shown in figure 1. In the same chamber, the e$^-$ beam setup is placed as shown in figure 1 (right). It consists of the crucible (for placing sample), shutter (initially used to collect the impurities if any are present and to stop them from depositing on the substrate during this process), and substrate holder attached with the rotating plate. The whole setup is covered with aluminium foil to prevent the deposition on the chamber walls. Figure \ref{fig:2} shows the schematic view of setup used for vacuum evaporation.

\section{Preparation Method}
Preparation method of LiF on self supporting silver foil consists of two steps. First step requires initial silver evaporation and second steps requires LiF deposition over deposited silver foil. The following steps describe the whole process efficiently.\newline

\begin{figure}
\begin{center}
\begin{tabular}{cc}

\includegraphics[width=65mm]{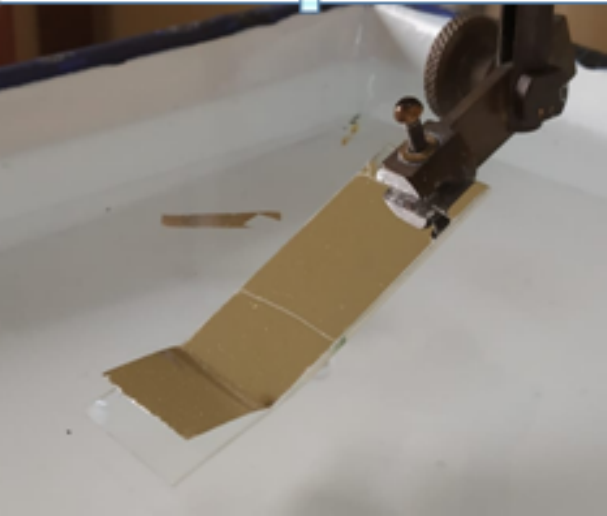}
\includegraphics[width=65mm]{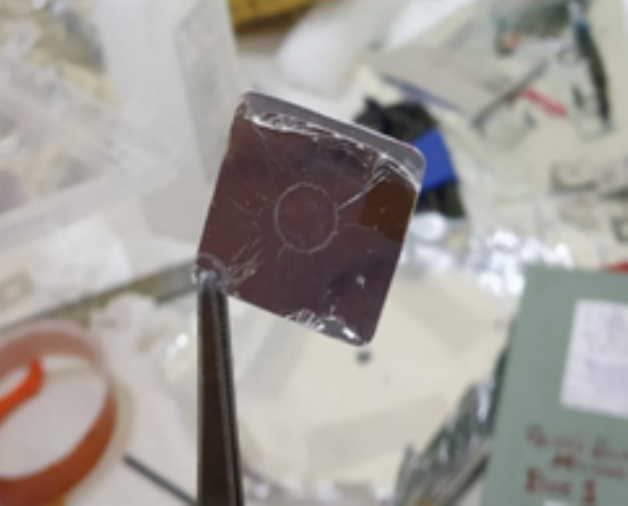} \\
\includegraphics[width=65mm]{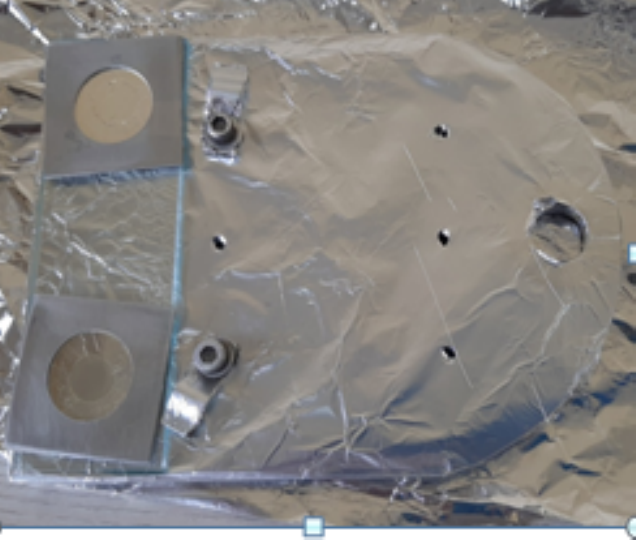}

\end{tabular}

\caption{(Left) Floating process of silver foil  (Right) Silver foil mount on Al frame (Middle) Deposited LiF on self supporting Ag foils mounted on substrate holder  }
\label{fig:3} 
\end{center}      
\end{figure}

(i) First, three glass slides of 76mm$\times$25mm are attached to substrate holder of rotating plate as shown in figure \ref{fig:1} (right). Glass substrates are cleaned with ethanol before attaching to the substrate holder. A thin layer of BaCl$_2$ is deposited through e$^-$ beam evaporation which will act as releasing agent. Then, pure silver (~99.99 $\%$) is deposited over the deposited BaCl$_2$ using e$^-$ beam evaporation. The deposition of Ag is started at 33 mA of current with 0.1 A$^0$/s initial deposition rate and gradually goes on increasing with increase of current up to 50 mA with final deposition rate of 2.2 A$^0$/s. After deposition of Ag, glass slide is immersed in distilled water. The inter layer of  BaCl$_2$ is dissolved in water and Ag foil floats on the surface of water. Ag foils were carefully picked up by Al frame in order to avoid the impact of surface tension of water. Figure \ref{fig:3} shows the floating process of Ag foil and silver foil mounted on Al frame of 8-10 mm diameter opening. Substrate and source was at a distance of 22.5 cm during whole deposition process. The substrate holder was rotating at speed of 5 rpm to maintain uniformity of deposition.
\ \\
(ii) Then, self supporting Ag foils on Al frame were fixed to another substrate holder (figure \ref{fig:1} left (e)) at 10.5 cm distance from the copper electrode. The reduction in distance of substrate holder from the sample placed in boat can be attributed to low vapour pressure of LiF sample. LiF was deposited successfully by resistive heating method. Ag foils mounted on Al frame is sandwiched with a 15 mm diameter frame for restricting deposition area of LiF.
\begin{figure}
\begin{center}

\begin{tabular}{cc}

\includegraphics[width=65mm]{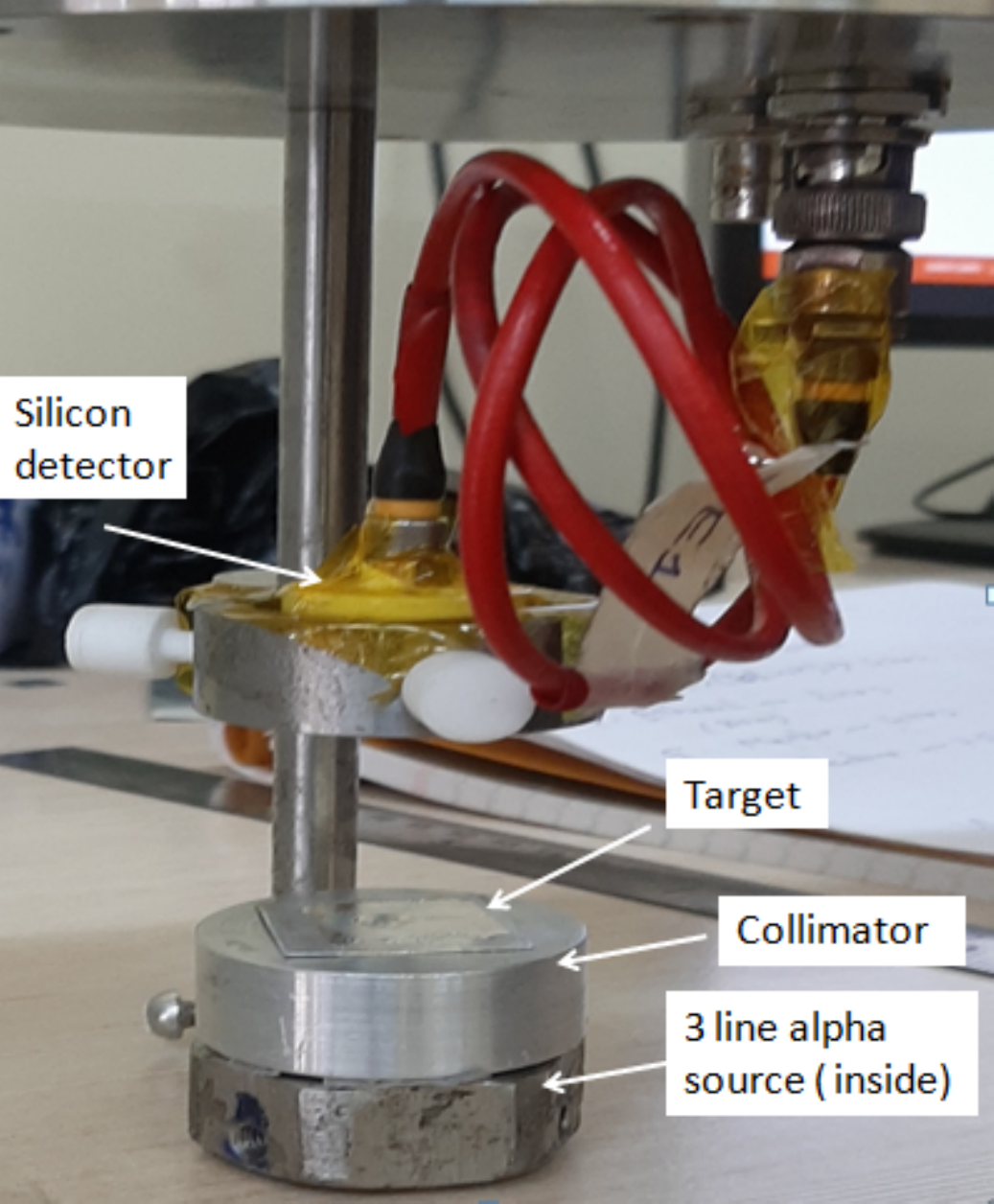}
\end{tabular}

\caption{Set-up for thickness measurement}

\label{fig:4} 
\end{center}      
\end{figure}

\begin{table}[]
\centering
\caption{Materials used and Properties}
\label{tab:1}       
\begin{tabular}{llll} \toprule
 Materials  & Melting point($^0$C) & Current & Mode of evaporation \\
 \toprule
 
  BaCl$_2$& 962 & 7-15 mA& ~~e$^-$ beam  \\
  \midrule
     Ag& 961.8 & 33-50 mA &~~e$^-$ beam   \\
      \midrule
     LiF&848.2& 60-65 A & ~~Resistive heating  \\
        \toprule

\end{tabular}
\end{table}
Evaporated LiF on Ag backing is shown in figure \ref{fig:3}.  Materials used for deposition, melting point, and current used for evaporation are tabulated in the table \ref{tab:1}.
\section{Analysis of prepared targets}
\begin{figure}[h!]
\begin{center}
\begin{tabular}{cc}
\includegraphics[width=70mm]{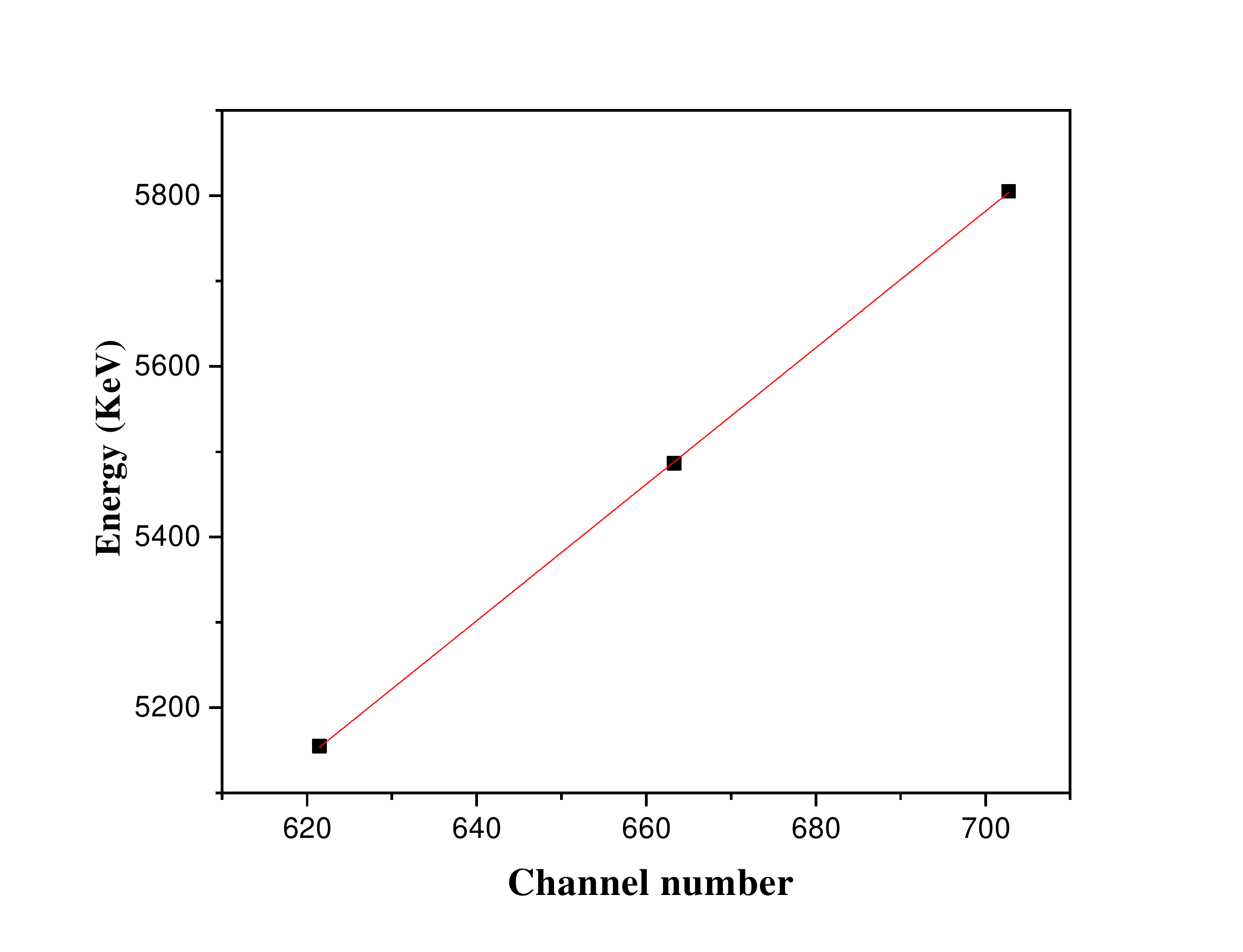} 
\includegraphics[width=70mm]{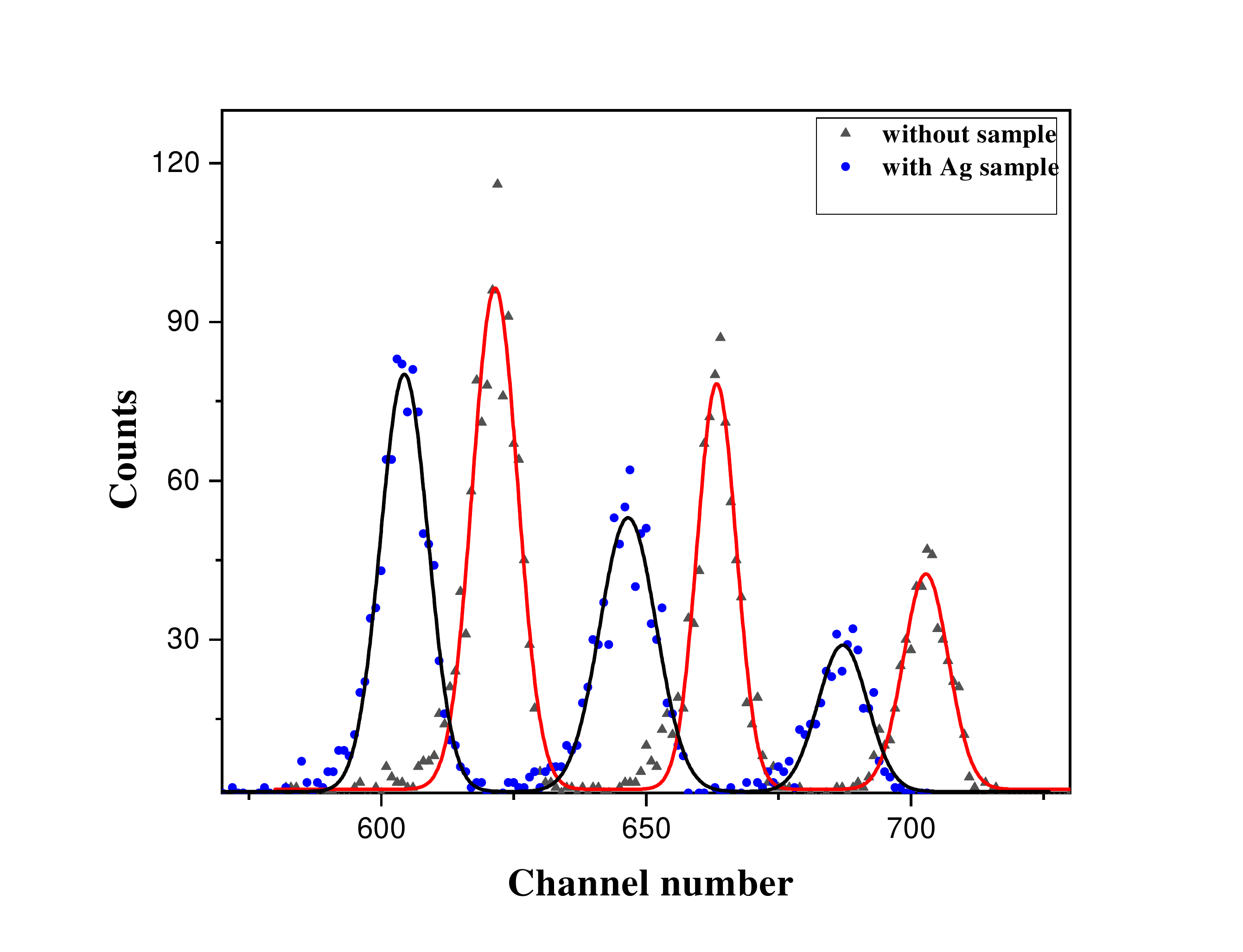}
\end{tabular}

\caption{(left) Calibration curve (right) Shift in peak  due to energy loss through Ag foil }
\label{fig:5} 
\end{center}      
\end{figure}

Thickness of prepared LiF targets on self supporting Ag backing is measured using three line alpha source ($^{239}$Pu, $^{241}$Am, and $^{244}$Cm).It is measured in two steps. First, thickness of Ag is measured using alpha energy loss.Then using known thickness of silver, energy loss through Ag can be calculated and  energy available after loss through Ag foil is used as initial incident energy for thickness calculation of LiF sample. Figure \ref{fig:4} shows the setup used for thickness measurement using alpha loss. Collimator shown in figure \ref{fig:4} is used to scan different position of samples for confirmation on uniformity of prepared targets. Silicon detector is used for detecting alpha particles after passing the sample. Setup shown in figure \ref{fig:4} was placed within a SS chamber with vacuum maintained at 10$^{-5}$ mbar. The three line alpha source was placed inside the groove before collimator. Energy associated with three line alpha source $^{239}$Pu, $^{241}$Am, and $^{244}$Cm are 5155 KeV, 5486 KeV and, 5805 KeV respectively. The detector needs to be calibrated properly to measure the thickness. This can be achieved by taking measurement without placing foil in front of alpha source.Figure \ref{fig:5} (left) shows the calibration curve of three line alpha source. The equation between energy of three line alpha and channel number is found to be 
\begin{equation}  \label{eq1}
Energy (KeV) = 183.77 + 7.99\times channel ~~ number
\end{equation}  
After placing the Ag foil in front of alpha source, the detector will detect the alpha particle with reduced energy. Figure \ref{fig:5} (right) shows the shift in peak of three line alpha energy with and without Ag foil respectively. Energy loss for three line energy can be calculated from the peak shift in figure \ref{fig:5} (right) using equation \ref{eq1}. Energy loss  ($\Delta$E) for three alpha energies are 136.93 KeV, 133.78 KeV, and 125.32 KeV. Stopping power for three alpha energy on Ag foils are 0.3361 KeV/($\mu$g/cm$^2$), 0.3241 KeV/($\mu$g/cm$^2$), and 0.3137  KeV/($\mu$g/cm$^2$). This stopping power is calculated using SRIM code \cite{7}. Thickness of the foil is calculated by

\begin{equation}  \label{eq2}
x = \frac{\Delta E}{-\frac{dE}{dx}}
\end{equation}  
where, x= thickness of foil in $\mu$g/cm$^2$, $\Delta$E = Energy loss through foil in KeV and -$\frac{dE}{dx}$ is the stopping power measured in KeV/ ($\mu$g/cm$^2$). Thickness of one of the Ag foil is 398.62 $\mu$g/cm$^2$ (0.38 $\mu$m) using equation \ref{eq2}. Uniformity of foil is checked by scanning the different position of foil through energy loss method with a collimator and variation in thickness of Ag foil is  between 5-10 \% from the average thickness of foil.
\newline
\ \\
\begin{table}[]
\centering
\caption{Thickness tabulation of LiF sample}
\label{tab:2}       

\begin{tabular}{ |m{2cm} |m{2cm}| m{2cm} |m{2.2cm} |m{2cm}|m{2cm}|} \hline
Energy loss in Ag (KeV) & Total energy loss in Ag+LiF (KeV)& Energy loss in LiF (KeV) & -$\frac{dE}{dx}$ of LiF (KeV/ ($\mu$g/cm$^2$))   & Thickness of LiF ($\mu$g/cm$^2$)& Avg. thickness of LiF ($\mu$g/cm$^2$) \\
 \hline
 
123.04 & 251.42  & 128.38& 0.7152 & 179.50  &\\
  \hline
  118.65   & 244.85 &126.20 &0.6845 &184.36 & 183.74 \\
      \hline
   114.84  &237.96&123.12 &0.6571 &  187.36 &\\
  \hline

\end{tabular}
\end{table}
However, LiF is evaporated over Ag foil. So the available energy of three lines alpha after loss of energy through Ag foil will be initial energy for LiF samples. So by subtracting the alpha energy loss through silver foil from initial energy of alpha source, the energy available for LiF sample are calculated. Figure \ref{fig:6} (left) shows the calibration curve for three alpha energies with channel number and the relation is found to be 
\begin{equation}  \label{eq3}
Energy (KeV) = 102.24 + 6.50\times channel ~~ number
\end{equation}  
Total energy loss for both silver and LiF samples are determined from the peak shift as shown in figure \ref{fig:6} (right). Thickness measurement of LiF sample for three alpha energy loss are tabulated in table \ref{tab:2}. Loss in energies through Ag in table \ref{tab:2} are calculated assuming the average thickness (0.35 $\mu$m) of silver foils.Thickness of LiF is calculated using the same relation (equation \ref{eq2}) as used for Ag. Uniformity of LiF sample is calculated using same procedure of collimator at different position. Maximum variation in thickness of LiF sample is with in 6 \% from the mean value.

\begin{figure}[h!]
\begin{center}
\begin{tabular}{cc}
\includegraphics[width=70mm]{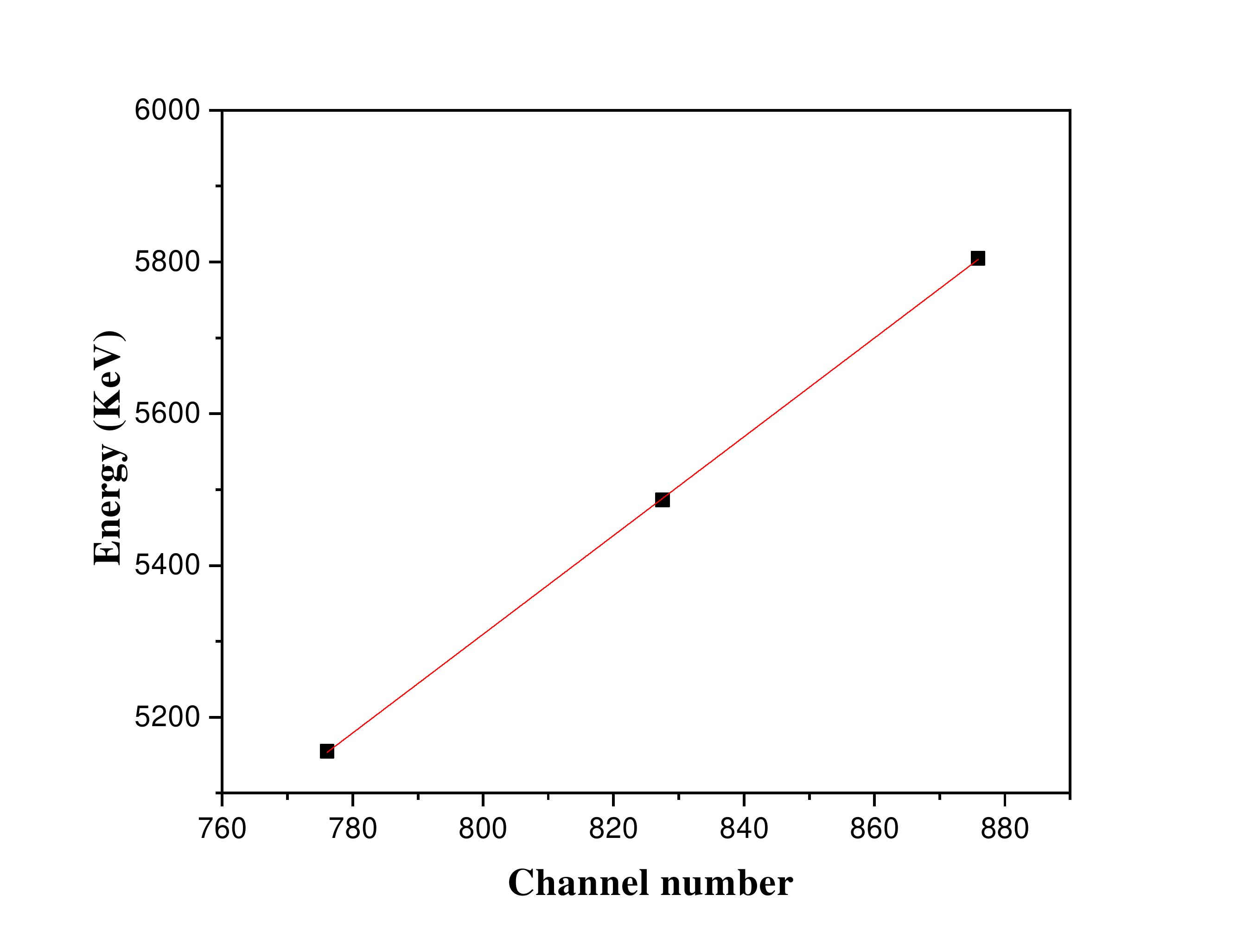} 
\includegraphics[width=70mm]{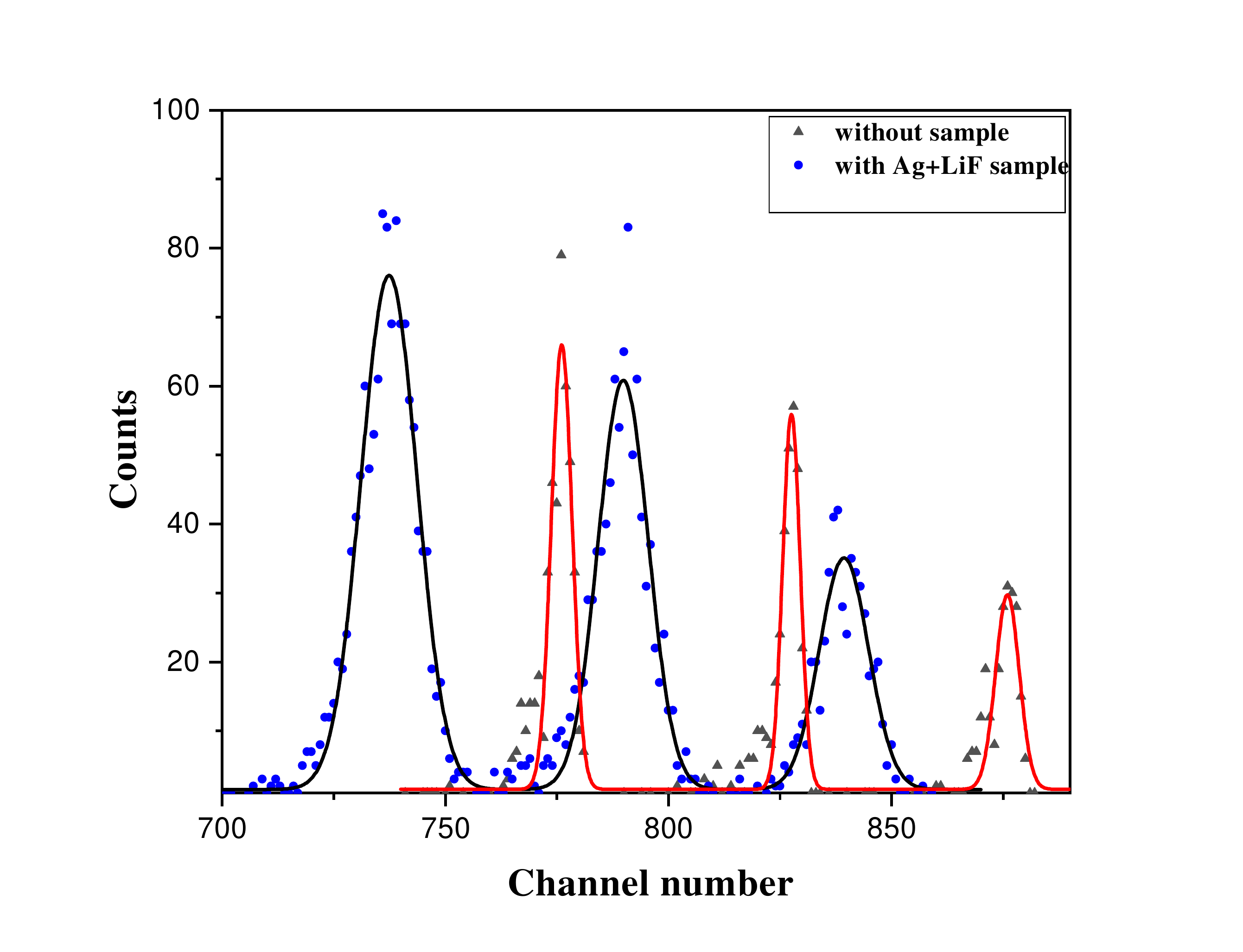}
\end{tabular}

\caption{(left) Calibration curve (right) Shift in peak  due to energy loss through Ag+LiF foil}
\label{fig:7} 
\end{center}      
\end{figure}

The presence of LiF samples on Ag backing are confirmed by X-ray photo electron spectroscopy (XPS). It is a surface sensitive technique used to identify the atoms present in sample surface. The XPS of samples were characterized using non-monochromatic Mg K$_\alpha$ x-ray source with energy of 1253.6 eV. The hemispherical e$^-$ analyser with radius of 150 mm and 20 eV pass energy were used to observe the ejected photo electrons. The XPS data were analysed  by XPSpeak41 software \cite{8}. The background is also subtracted by Shirley method using XPSpeak41. The figure \ref{fig:7} shows the XPS spectra of LiF sample. The spectra shows peaks at 29.58 eV  and 684.62 eV  denoting F 2s \cite{9} and F 1s \cite {10} respectively. The peak at 55.54 eV denotes Li 1s state \cite{11}. XPS analysis confirms the presence of 47.43 $\%$ of Li and 52.56 $\%$ of F atoms in the deposited targets.

\begin{figure}[h!]
\begin{center}
\begin{tabular}{cc}
\includegraphics[width=80mm]{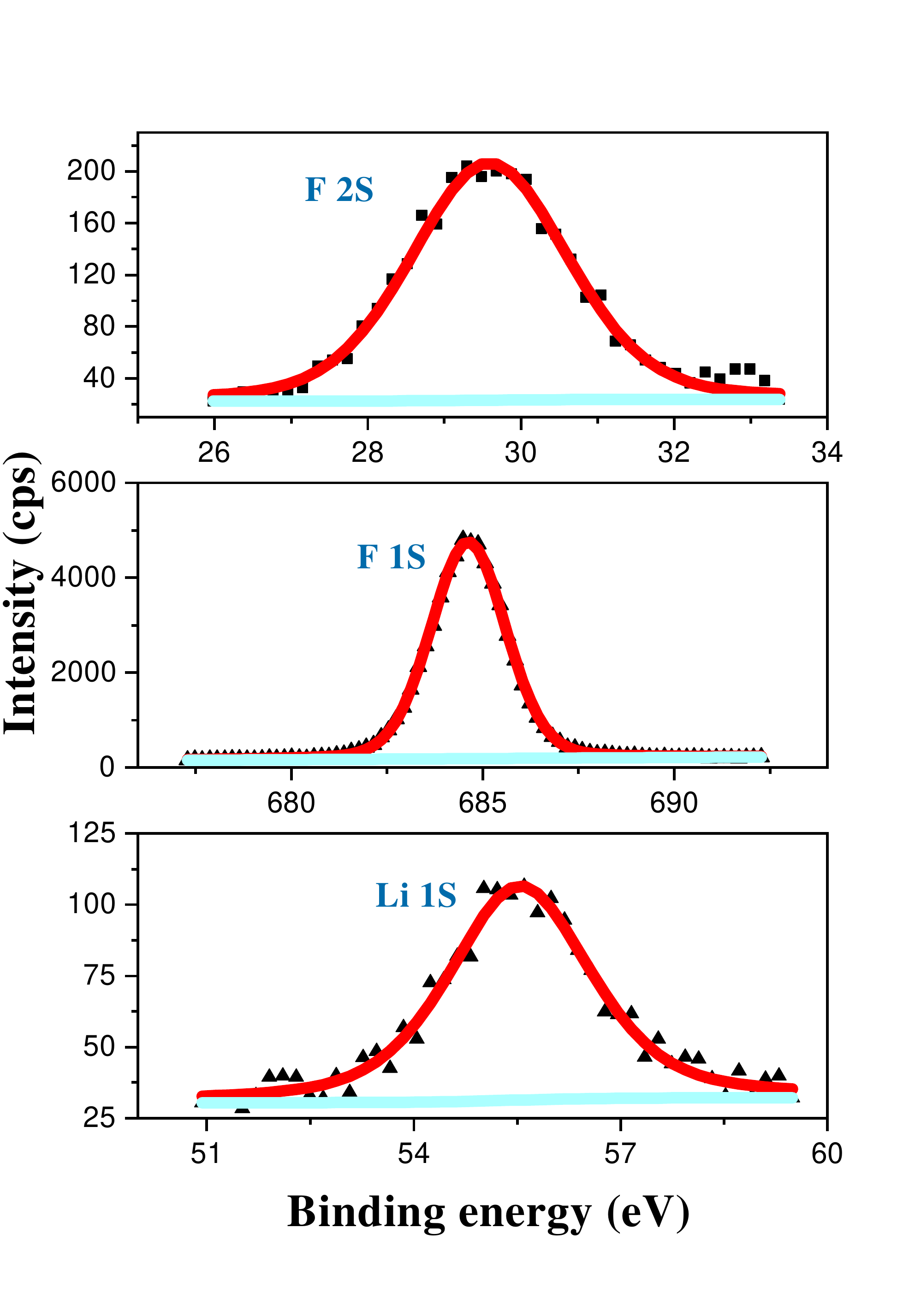} 
\end{tabular}

\caption{XPS spectra of LiF deposited on Ag backing}
\label{fig:6} 
\end{center}      
\end{figure}
\section{Conclusion}
In this paper, a smooth and detailed procedure is discussed for fabrication of LiF target on self-supporting Ag backing. Good number of LiF targets are fabricated successfully on self-supporting Ag backing. The thickness of one of the target of LiF is found to be 183.74 $\mu$g/cm$^2$ by observing loss of energy of three line alpha source. However, the thickness can be tuned by monitoring the current and evaporation time. Maximum non-uniformity observed is within 6 $\%$. So good uniformity is achieved during deposition. This targets are intended to be used for mostly alpha and proton induced reaction especially suitable for nuclear astrophysics experiments. The high melting point of LiF is also suitable for use in high current accelerator facility. XPS spectra confirms the presence of Li and F atom on the surface of sample. The atomic concentration observed from XPS analysis is 47.43 $\%$ of Li and 52.56 $\%$ of F.   

\section{Acknowledgement}
The authors would like thank Prof. Supratic Chakraborty, Prof. Satyazit Hazra, Ms. Mousri Paul, and Mr. Subhankar Mondal of Surface Physics and Material Science division,SINP,Kolkata for allowing us to use XPS facility for material characterization. 

 \bibliographystyle{elsarticle-num} 
 \bibliography{cas-refs}





\end{document}